\documentclass[pre,floatfix,twocolumn,a4paper,nofootinbib,amssymb,
               amsmath,showpacs,superscriptaddress]{revtex4}
\usepackage{graphicx}

\newcommand {\be} {\begin{equation}}
\newcommand {\ee} {\end{equation}}
\newcommand {\Be}{\begin{eqnarray*}}
\newcommand {\Ee} {\end{eqnarray*}}
\newcommand {\bey} {\begin{eqnarray}}
\newcommand {\eey} {\end{eqnarray}}
\newcommand{\bit}{\begin{itemize}}      
\newcommand{\eit}{\end{itemize}}
\newcommand{\bfl}{\begin{flusleft}}
\newcommand{\efl}{\end{flusleft}}
\newcommand{\bfr}{\begin{flushright}}
\newcommand{\bc}{\begin{center}}
\newcommand{\ec}{\end{center}}
\newcommand{\ben}{\begin{enumerate}}    
\newcommand{\een}{\end{enumerate}}

\newcommand{\comment}[1]{}
\newcommand{\G}{\Gamma}
\newcommand{\g}{\gamma}
\newcommand{\s}{\Theta}
\newcommand{\av}[1]{\left\langle #1\right\rangle}

\begin{document} 

\title{Desynchronization in diluted neural networks} 

\author{R\"udiger Zillmer}
\email{zillmer@fi.infn.it}
\affiliation{INFN Sez. Firenze,
via Sansone, 1 - I-50019 Sesto Fiorentino, Italy}
\author{Roberto Livi}
\email{livi@fi.infn.it}
\affiliation{Dipartimento di Fisica, Universit\'a di Firenze,
via Sansone, 1 - I-50019 Sesto Fiorentino, Italy}
\affiliation{Sezione INFN, Unita' INFM e Centro Interdipartimentale per lo Studio delle Dinamiche
Complesse, via Sansone, 1 - I-50019 Sesto Fiorentino, Italy}
\author{Antonio Politi}
\email{antonio.politi@isc.cnr.it}
\affiliation{Istituto dei Sistemi Complessi, CNR, 
CNR, via Madonna del Piano 10, I-50019 Sesto Fiorentino, Italy}
\affiliation{Centro Interdipartimentale per lo Studio delle Dinamiche
Complesse, via Sansone, 1 - I-50019 Sesto Fiorentino, Italy}
\author{Alessandro Torcini}
\email{alessandro.torcini@isc.cnr.it}
\affiliation{Istituto dei Sistemi Complessi, CNR, 
CNR, via Madonna del Piano 10, I-50019 Sesto Fiorentino, Italy}
\affiliation{Centro Interdipartimentale per lo Studio delle Dinamiche
Complesse, via Sansone, 1 - I-50019 Sesto Fiorentino, Italy}


\begin{abstract}

The dynamical behaviour of a weakly diluted fully-inhibitory network of 
pulse-coupled spiking neurons is investigated. Upon increasing the 
coupling strength, a transition from regular to stochastic-like regime is
observed. In the weak-coupling phase, a periodic dynamics is rapidly approached,
with all neurons firing with the same rate and mutually phase-locked. The
strong-coupling phase is characterized by an irregular pattern, even though the 
maximum Lyapunov exponent is negative. The paradox is solved by drawing an
analogy with the phenomenon of ``stable chaos'', i.e. by observing that the 
stochastic-like behaviour is ``limited" to a an exponentially long 
(with the system size) transient. Remarkably, the transient dynamics turns out
to be stationary.
\end{abstract} 



\maketitle


\section{Introduction}
\label{one}

During the last few years it has become increasingly clear that understanding 
the behaviour of many different systems passes through the comprehension of
the dynamics of complex networks \cite{network}. This is, for instance, the case 
of metabolic systems, genetic networks, the immune response system and 
neurobiological structures \cite{network}. A particular challenge is represented
by the need of unravelling the mutual connections between network structure and
dynamical properties. Very little is in fact known about the expected
classes of behaviour and their stability properties even in systems of globally 
coupled identical oscillators. For this reason and under the assumption of a
structural stability of the possible scenarios, it is, therefore,
instructive to investigate simple models, such as diluted neural networks of
pulse--coupled neurons. Since it seems that inhibition plays a major role in
determining the dynamics of single neo-cortical piramidal neuron \cite{rudolph}
as well as of cortical networks \cite{brunel-wang}, we have chosen to examine a
network  of inhibitory coupled leaky integrate--and--fire neurons. More 
precisely, we consider the model proposed in Ref.~\cite{Jin:2002}, where Jin, 
under fairly general conditions, investigated analytically the convergence 
towards a periodic pattern. In this paper we study the diluted version of this 
model, showing that even though the dynamics is characterized by a negative
maximum Lyapunov exponent \cite{foot0}, irregular and exponentially-long
transients are typically observed for a sufficiently strong coupling strength.
This phenomenon is somehow analogous to what observed in diluted networks of
pulse-coupled oscillators with delay \cite{Zumdieck-Timme:2004}, although the
transient therein reported are chaotic in the typical sense of the word. More
precisely, we find that for small coupling amplitudes the dynamics converges,
after a short transient, towards a synchronized state with all neurons firing
with the same rate, but with their phases approximately uniformly distributed.
This can be understood by first referring to a homogeneous network of globally
coupled neurons. In that context, the mean field is found to induce an
effective repulsive interaction between the neurons; as a result the asymptotic
regime is characterized  by an evenly spaced sequence of spikes which is known in
the literature as a {\it splay state} \cite{splay}. As a result of random
dilution, one can imagine that inhomogeneities in the mutual interactions arise
which in turn lead to small nonuniformities in the interspike intervals.

On the other hand, for sufficiently large coupling amplitudes, stochastic-like
transients are observed, whose duration is exponentially long with the network
size. Since various indicators show that this regime is stationary, it is
logical to conclude that in infinitely large networks it represents a
perfectly legitimate thermodynamic phase. This is analogous to the active phase
in directed percolation, whose life-time is finite in finite systems. Even more
stringent is the analogy with ``stable chaos", a kind of irregular behaviour
discovered in coupled map lattices\cite{stable} and characterized by negative
Lyapunov exponents. Since it is believed that such a pseudo-chaotic behaviour
is sustained by discontinuities of the mapping rule\cite{Ginelli}, it is natural
to expect this to be true also in the present case. Quite consistently, we
observe that in the presence of disorder, where neurons are no longer
equivalent to one another, changes in the firing order are accompanied by
discontinuities in the evolution rule.

Altogether, the transition manifests itself as a collective de-synchronization
phenomenon. Unfortunately, a ``microscopic" linear--stability analysis does not
allow identifying the threshold, since all trajectories are asymptotically
stable in both regimes. Moreover, direct numerical simulations are not very
effective either, due to the difficulty of simulating large networks, so that
the study of the critical behaviour is an even more difficult task. However, the
evidence of two distinct phases is rather convincing and, furthermore, the
introduction of a suitable space-time representation suggests a true analogy
with directed percolation that will be worth exploring in more detail.

In Sec.~\ref{two}, we introduce the model and the variables, while the
homogeneous case of fully coupled networks is analytically investigated
in Sec.~\ref{three}, where we also introduce a one-dimensional description of
the dynamics that applies exactly in the thermodynamic limit.
In Sec.~\ref{four} the transient dynamics of weakly diluted networks is
discussed, both by determining the transient length and analysing its
stationarity properties. The resulting two-phase scenario is then summarized
in Sec.~\ref{five}, where we also present an effective stochastic description.
Further comments about open questions and indications for future studies
are finally presented in the conclusions.


\section{The model} 
\label{two}

In this paper we investigate a system of $N$ leaky integrate-and-fire (LIF) 
neurons, analogous to the model discussed in Ref.~\cite{Jin:2002}. The state of 
the $i$-th neuron is fully determined by the membrane potential $V_i(\tilde t)$ 
and obeys the differential equation
\be
\label{eq:model1}
\tau\dot V_{i}=C-V_{i}-\tau(V_i+W)\, \sum_{j=1}^{N}\sum_{m} g_{ij} 
\,
\delta(\tilde t-\tilde t_j^{(m)})\ ,
\ee
where $\tau$ is the membrane time constant, $C$ is the suprathreshold input
current (referred to a unitary membrane resistance), and $W$ is the reversal
potential. Whenever the potential $V_j(\tilde t)$ reaches the threshold value
$\Theta$, it is reset to $R<\Theta$, and a spike is sent to and
instantaneously received by all connected neurons at time $\tilde t_j^{(m)}$
(the superscript $m$ enumerates the firing events of the $j$-th neuron). The
net result of a received spike is that the membrane potential of the $i$-th
receiving neuron is decreased according to the transformation
\be
\label{eq:inhib}
V_i' + W = (V_i + W) \exp(-g_{ij})    .
\ee
As a consequence of the inhibitory connections, the potential $V_i'$ can go
below the reset value $R$, but Eq.~(\ref{eq:inhib}) shows that $-W$
is a true lower bound. However, for small coupling values and in the absence 
of clustering phenomena (as in the present manuscript), it turns out that $V_i$
is essentially bounded between $R$ and $\Theta$. The last ingredient defining the
system dynamics is the connectivity matrix  $g_{ij}$. Following the recent
literature on randomly connected directed networks
\cite{Zumdieck-Timme:2004,Gerstner-Kistler:2002}, the coupling strength
is scaled to the connectivity of the receiving neuron,
\be
\label{eq:norm}
g_{ij} =
  \begin{cases}
   G/\ell_i \,\,\,\,\, , \,\,\,\,\, {\rm if}\,\,\,\, i \,\,\,\, {\rm and}\,\,\,\, j \,\,\,\, 
   {\rm are} \,\,\,\, {\rm coupled}
\\
   0 \,\,\,\,\, , \,\,\,\,\, {\rm otherwise} 
\ 
    \end{cases}\ ;
\ee
where $G$ is the coupling constant and $\ell_i$ is the number of incoming links
to neuron $i\,$. In other words, we consider the simplest type of disorder,
determined just by the presence/absence of links between neurons (notice that 
self-interactions are excluded, i.e. $g_{ii} = 0$). As we are interested in
studying networks with a given fraction $r_m$ of missing links, there are in
principle different ways of doing that: ({\it i}) each link is cut with a
probability $r_m$; ({\it ii}) the total number $N_m$ of cut links is fixed
deterministically ($N_m = r_m N(N-1)$); ({\it iii}) the number of cut
links per neuron is fixed deterministically. Since preliminary simulations
performed according to the three philosophies have given qualitatively similar
results, we have eventually decided to restrict our quantitative analyses to the
second approach. Moreover, since we aim at understanding the possibly
qualitative changes induced by disorder in the neural network dynamics, we
limit ourselves to analysing the case of weak disorder. This is done by choosing
small values for the fraction $r_m$ (typically $r_m = 0.05$). 

Most of the parameter values are set according to the current literature 
(see e.g., \cite{Brunel-Hakim:1999,Jin:2002}), namely: $\tau=20\, \text{msec}$,
$C=-45\,\text{mV}$, $W=63\,\text{mV}$, $\Theta=-52\,\text{mV}$, and
$R=-59\,\text{mV}$. As a matter of fact, the only free parameters of the model
are the fraction $r_m$ and the coupling constant $G$, which tunes the strength 
of the inhibitory coupling. 

The dynamical equation (\ref{eq:model1}) can be recasted into a simpler form by
introducing the dimensionless parameters 
\begin{gather*}
t=\tilde t/\tau\,,\, v_i = (V_i-R)/(\s-R)\,,\\
  c= (C-R)/(\s-R)\,,\, w=(W+R)/(\s-R) \,.
\end{gather*}
This amounts to rescaling $\Theta$ and $R$ to the values 1 and 0, respectively.
Moreover, for the above choice of parameter values, $c=2$ and $w=4/7$.

As a result, Eq.~\eqref{eq:model1} reads
\be
\label{eq:model2}
\dot{v}_{i}=c-v_{i}-(v_i+w)\,
  \sum_{j=1}^{N}\sum_{m}g_{ij}
  \,\delta(t-t_j^{(m)})\,\,\, .\,\, 
\ee
Since $w >0$, the coupling is fully inhibitory, i.e., an incoming spike lowers
the potential of the receiving neuron and hence the coupling inhibits firing. 

At variance with the original model \cite{Jin:2002}, where both the threshold 
currents $C_i$ and the coupling constants $g_{ij}$ are assumed to be randomly 
distributed, here the only source of disorder is the presence/absence of
inhibitory connections. This choice is dictated by our interest in relatively
simple structures to better understand the possible scenarios. However, the
most important difference concerns the coupling constant: no dependence on the 
system size is postulated in Ref.~\cite{Jin:2002}, while an inverse 
proportionality to the number of incoming connections is assumed here.

This latter normalization is more suited to investigate the large $N$-limit,
since it guarantees that the coupling strength remains comparable to the
amplitude of the force field ($c-v$). 

By following the approach of Ref.~\cite{Jin:2002}, we map the original model
onto a discrete-time map. Let $v_i(n)$ denote the membrane potential of the
$i$-th neuron immediately after the $n$-th spike. Until the next spike is 
emitted, the network evolution corresponds to an exponential relaxation of each
$v_i$ towards $c$. One can easily infer from (\ref{eq:model2}) that the time
interval $\Delta t_i$ needed by the $i$-th neuron to reach the firing threshold
$v_i=1$ is
\be
\label{eq:map}
\Delta t_i(n) := \ln\G_i(n)\quad\text{with}\quad\G_i(n)=\frac{c-v_i(n)}{c-1}\ .
\ee
Let now $k(n)$ denote the label of the neuron characterized by the shortest 
time, i.e.,
\be
\label{eq:map1}
\G_{k(n)}=\min_i\{\G_i\}  \, .
\ee
By now including the effect of the next spike, the network dynamics can be
be transformed into a discrete map for the quantities $\G_i(n)$. In the
large $N$ (and, accordingly, $\ell_i$) limit, one obtains
\begin{subequations}
\label{eq:map2}
\begin{gather}
\G_k(n+1)=\frac{c}{c-1}\ ,\\
  \G_i(n+1)= g_{ik}\,a+( 1 -g_{ik})
  \frac{\G_i(n)}{\G_k(n)}\quad\text{for}\ i\ne k\ ,
\end{gather}
\end{subequations}
where
\be\label{eq:param}
a :=\frac{c+w}{c-1}\ .
\ee
The process can be iterated by finding the minimum among the $\G_i(n+1)$
and so on. Therefore, $\Delta t_{k(n)}(n)$ represents the time interval between 
the $n$-th and the $(n+1)$-st spike; following the standard notation, it will be
denoted with $t_{\rm ISI}(n)$. On the other hand $T(m)$ denotes the time elapsed
between the $(m+1)$-st spike and the previous spike emitted by the {\it same}
neuron.


\section{The homogeneous case}
\label{three}
In this section we discuss the dynamics of homogeneous networks, i.e.,
$g_{ij} =G/\ell_i$ (for $i\ne j$) and $\ell_i=N$. In the absence of any disorder
(globally coupled identical neurons), any initial ordering of the neuron spikes
will persist at all times, because it cannot be changed by the dynamical rule.

Accordingly, after a short transient, the trajectory converges towards a
periodic orbit characterized by equispaced spikes, $t_{\rm ISI}(n)= T/N $ and
$T(m) = T$ for all $n$, $m$. By formally interpreting the separation between the
spike time $t^{(m)}_j$ of the $j$th neuron and the preceding spike-time
$t^{(n)}_i$ of some reference neuron as a phase variable, one can summarize the
scenario by stating that the single-neuron phases repel each other until an
equilibrium state, characterized by a maximal and uniform separation, is
attained. In the literature, this alignment is called splay state and it has
been found in various models of globally coupled oscillators \cite{splay}. 
Such a state can be viewed as the opposite situation of a fully synchronized
state, where all oscillators share the same phase. It is curious that this
regime is attained in the presence of inhibitory coupling, since this property
is usually associated with the propensity to synchronize instead. In fact,
splay states have been found in globally coupled LIF neurons in the presence of
excitatory coupling with spikes of finite width \cite{vvres,mohanty}.
 
\subsection{Stationary solution}
Since the ordering of the neuron potentials, $v_i$, does not change in
time, one is free to label the spiking neuron in such a way that
$k(n+1) = k(n)+1 \mod N $. It is also convenient to choose a ``moving"
frame, namely $j = i-n$, since the label of the spiking neuron remains constant
and can thereby be set equal to 1, without loss of generality. The resulting
map reads 
\be
\label{eq:fixp1}
\G_{j-1}(n+1)=\frac{a\,G}{N}+\left(1-\frac{G}{N}\right)
  \frac{\G_j(n)}{\G_1(n)}\ ,
\ee
with the boundary condition $\G_N=\tilde c :=c/(c-1)$. This equation 
admits a stationary solution, which corresponds to a regime of evenly spaced
spikes, i.e. a  splay state~\cite{splay}. 

The fixed point of \eqref{eq:fixp1} is obtained by solving the recursive 
relation
\[
  \tilde\G_{j-1}=\frac{a\,G}{N}+\left(1-\frac{T+G}{N}\right)\tilde\G_j\ ,
\]
where $\tilde\G_1$ is approximated with $1+T/N\,$. In fact, from 
the very definition, one has $\ln \tilde \G_1 = t_{\rm ISI}= T/N$, so that
$T$ is the (constant) single neuron interspike interval. By iterating backward
the above equation from the initial condition $\tilde\G_N=\tilde c$, one obtains
\be
\label{eq:fixp2}
\tilde\G_{N-n}=\frac{a\,G}{T+G}\left[1-e^{-(T
  +G)n/N}\right]+\tilde c\,e^{-(T+G)n/N}\ .
\ee
The value of $T$ can be finally determined by imposing the condition
\be
\label{eq:defT}
\tilde\G_1=1+T/N\ .
\ee
By neglecting first order corrections in $1/N$, one obtains,
\be
\label{eq:fixp3}
1=\frac{a\,G}{T+G}\left[1-e^{-(T+G)}\right]+\tilde c\,e^{-(T+G)}\ .
\ee

\subsection{Linear Stability}
Let us consider the evolution of an infinitesimal perturbation $\g_j(n)\,$ of 
the fixed point $\tilde\Gamma_j\,$. From the linearization of Eq.~(\ref{eq:fixp1})
it follows that $\g_j(n)$ satisfies the recursive relation
\be
\label{eq:fixl1}
\g_{j-1}(n+1)=\left(1-\frac{T+G}{N}\right)\left[\g_j(n)-
  \left(1-\frac{T}{N}\right)\tilde\G_j\g_1(n)\right]\, ;
\ee
where higher order corrections in $1/N$ are neglected. Because of the 
reset $\tilde\G_N=c/(c-1)$, it is natural to impose the boundary condition 
$\g_N=0\,$. By further setting $\g_{j}(n+1)=\mu\,\g_{j}(n)$, one
obtains the eigenvalue equation,
\be\label{eq:fixl2}
\g_j-\left(1-\frac{T}{N}\right)\tilde\G_j\g_1-\tilde\mu\,\g_{j-1}=0\ ,
\ee
where
\[
  \tilde\mu := \frac{\mu}{1-(T+G)/N}\ .
\]
By iterating the above equation and imposing the boundary condition, we find
that $\tilde\mu\,$ satisfies the equation
\[
  \tilde\mu^{N-1}+\left(1-\frac{T}{N}\right)\sum_{j=0}^{N-2}
  \tilde\G_{N-j}\,\tilde\mu^j=0\ .
\]
By inserting the expression of $\tilde\G_{N-j}$ (see Eq.~\eqref{eq:fixp2})
and using the condition \eqref{eq:fixp3}, we find that for $N\gg 1$
\be
\label{eq:fixl3}
\tilde\mu=-1-\frac{\ln \tilde c}{N}\ ;
\ee
which yields the expression of the Lyapunov exponent of the discrete map
\be
\label{eq:fixl4}
\lambda=\ln|\mu|=-\frac{1}{N}\left[T+G-\ln \tilde c \right]\ .
\ee
The Lyapunov exponent of the original system is finally obtained by
rescaling time with the interspike interval $T/N$,
\be
\label{eq:fixl5}
\Lambda=\frac{N}{T}\ln|\mu| = -1 -\frac{G -\ln \tilde c}{T} \ .
\ee

\begin{figure}[h]
\includegraphics[clip,width=6.5cm]{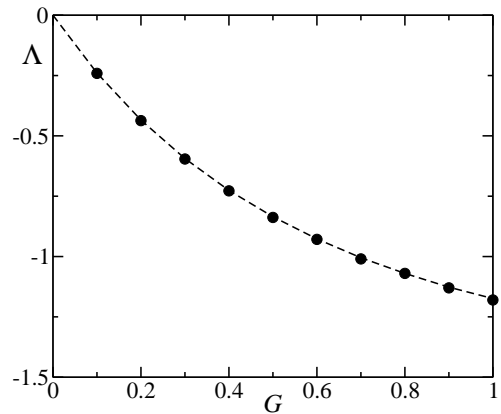}
\caption{The maximal Lyapunov exponent $\Lambda$ corresponding to \eqref{eq:fixp1} 
for $r_{\rm p}=0, c=2, w=4/7$ versus the coupling $G$. The dashed line refers
to the analytical expression \eqref{eq:fixl5} and the symbols to numerical
estimates.}
\label{fig:fixl1}
\end{figure}

In Fig.~\ref{fig:fixl1} we plot $\Lambda$ and compare it with numerical results
for different values of the coupling $G\,$. It is worth recalling that
$T\equiv T(G)$ as from Eq.~(\ref{eq:fixp3}).

\subsection{Continuous approach}

In the large $N$ limit, the discrete time ($n$) and ``space'' ($j$) variables can
be transformed into continuous ones by introducing
\[
  x =j/N\,,\ x\in [0,1]\,,\ \tau= n/N\ .
\]
As a result, by neglecting $1/N^2$ terms, Eq.~\eqref{eq:fixp1} writes as
\be\label{eq:12}
-\frac{\partial \G}{\partial x}+\frac {\partial \G}{\partial\tau}=
  -(T+G)\G+a\,G\ ;
\ee
where we have made use of Eq.~(\ref{eq:defT}). The boundary conditions read now
$\G(0,\tau)=1$ and $\G(1,\tau)=\tilde c\,$. The stationary solution corresponds
to the fixed point of \eqref{eq:fixp1} and reads
\begin{gather}
\label{eq:13}
\tilde\G(x)=\left(1-\frac{a\,G}{T+G}\right) e^{(T+G)x}
  +\frac{a\,G}{T+G}\ .
\end{gather}
The period $T$ is defined by imposing the boundary condition at $x=1$, i.e.,
\[
\left(1-\frac{a\,G}{T+G}\right) e^{(T+G)}
  +\frac{a\,G}{T+G}=\tilde c\ .
\]
This result is consistent with Eq.~(\ref{eq:fixp3}). Thus, we see that the
evolution equation of the globally coupled homogeneous network can be reduced
to a 1+1 dimensional partial differential equation. We shall see later that
this analogy proves fruitful to interpret the phase transition discussed in the
following sections.


\section{Transient dynamics}
\label{four}

From now on we study the dynamics of model \eqref{eq:model2} when a small fraction
$r_m$ of links is removed. It turns out that in this case a much more
interesting and complex dynamics may emerge when the coupling strength is
increased, even though the Lyapunov exponent remains negative. However, at
variance with the previous section, here we can rely on numerical simulations
only.

\begin{figure}[h]
\includegraphics[clip,width=6.5cm]{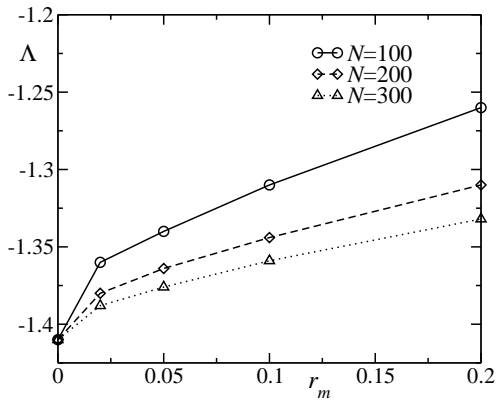}
\caption{The maximal Lyapunov exponent $\Lambda$ 
for different values of the cut ratio $r_{m}$ and for three values
of the system size. The data refer to a coupling $G =2$.}
\label{fig:lambda1}
\end{figure}

We expect that the weak amount of disorder induced by the random pruning of
directed links reduces the coupling strength among the neurons and
correspondingly increases the value of $\Lambda$, with respect to the fully
coupled case. The results plotted in Fig.~\ref{fig:lambda1} show that this is
indeed the case, as the Lyapunov exponent is found to increase with the pruning 
ratio $r_{m}\,$. Nevertheless, $\Lambda$ remains negative up to $r_{m}=0.2$.
Moreover, numerical simulations indicate that $\Lambda$ remains finite in
the thermodynamic limit $N \to \infty$.

As a result, sooner or later the dynamics must converge towards a stable
periodic orbit. In Ref.~\cite{Jin:2002}, Jin has derived an upper bound for the
transient length in the case of generic disorder. More precisely, he finds that
the number of spikes preceding a periodic pattern is
\be 
\label{eq:jin1}
P \approx N^q + (q - 1)\ ;
\ee
where, in the large $N$ limit, and with reference to our notations,
\begin{gather*} \label{eq:jin2}
q = -\ln \left( \Delta/8 a^3\right)/g_{min} + 1\ ,\\
g_{min} = \min_{i,j}\{g_{i,j}\}\ ,\ \text{and finally,}\\
\Delta= \min_{n=1,\dots,\infty} \min_{j \ne k(n)}\left[
\Gamma_j(n) - \Gamma_{k(n)}(n) \right] \ .
\end{gather*}
Although the $N$ dependence expressed by Eq.~(\ref{eq:jin1}) superficially looks 
like a power law, it is in fact much faster. First of all, if the coupling
strength scales as $1/N$ (as it is assumed here), the exponent $q$ grows 
linearly with $N$. Incidentally, this is true also in the fully coupled regime,
where we have seen that the transient is, in reality, much shorter. A second
reason is that in systems with some pruning (like here), $g_{min}$ is equal to
0 and the exponent $q$ would be actually equal to infinity. Finally, $\Delta$, 
which is the minimal distance between the membrane potential of the two neurons
that are closest to the firing threshold, cannot be larger than $1/N$. This
leads to an additional logarithmic growth of the exponent $q$ with $N$. 
Accordingly, we conclude that the analytic estimate contained in 
Ref.~\cite{Jin:2002} applied to our setup is much too large an overestimation of
the transient length to help understanding when and whether a phase
transition can occur. 

The transient duration $t_{\rm tr}$ is determined as the smallest time
for which the actual configuration of the membrane potential is
$\varepsilon$-close to a previous state,
$$
t_{\rm tr} = \min \{ t |
 \max_{\,i}|v_i(t+T)-v_i(t)|<\varepsilon\, ,\}
$$
where $1\le i \le N$ and $1 \le T \le t-1$. The choice of the threshold is not
crucial, since after the maximal distance is on the order of $1/N$, it starts
converging exponentially fast to 0. Accordingly, by this procedure, we do not
only determine the transient length but also the periodicity $T$ of the
asymptotic solution.

The dynamics (\ref{eq:model2}) is usually simulated by starting from a random
initial condition with the values of the membrane potentials $v_i(0)$ uniformly
distributed in the interval $[-w,1]$. In the following we present a detailed
analysis carried out for networks with $r_{m}=0.05$. 

\begin{figure}[h] 
\includegraphics[width=0.4\textwidth,clip=true]{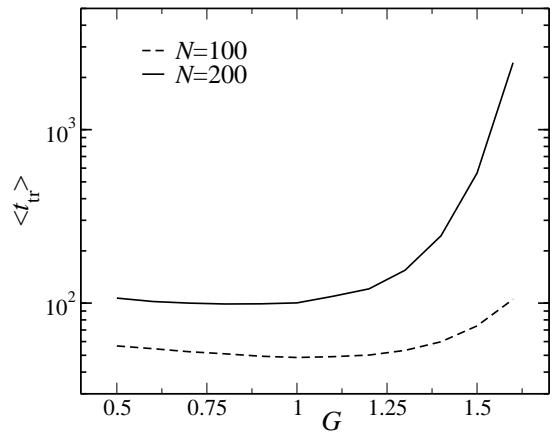}
\caption{Average transient length 
$\av{t_{\rm tr}}$ {\em vs.} $G$ for two system sizes.}\label{fig:trans}
\end{figure}

In order to deal with a more reliable quantity, we compute the average length
of the transient $\langle t_{\rm tr}\rangle$ (here and in the following, angular
brackets denote an average over both realizations of the network and different
initial conditions of the membrane potentials). In Fig.~\ref{fig:trans}, we have
plotted the average transient versus $G$ for two different values of the system
size ($N=100$ and 200). For $G\alt 1$, $\av{t_{\rm tr}}$ slightly decreases for
increasing coupling strength. This is simular to what observed in the fully
coupled network over the entire range of variation of $G$ \cite{Jin:2002}.
On the other hand, by further increasing the coupling strength
$\av{t_{\rm tr}}$ exhibits a sudden increase. More important is that the rate of
increase is significantly larger when $N$ doubles. Altogether this data suggests
the existence of two distinct phases approximately corresponding to 
$G$ smaller and larger than 1.

\begin{figure}[h]
\includegraphics[width=0.4\textwidth,clip=true]{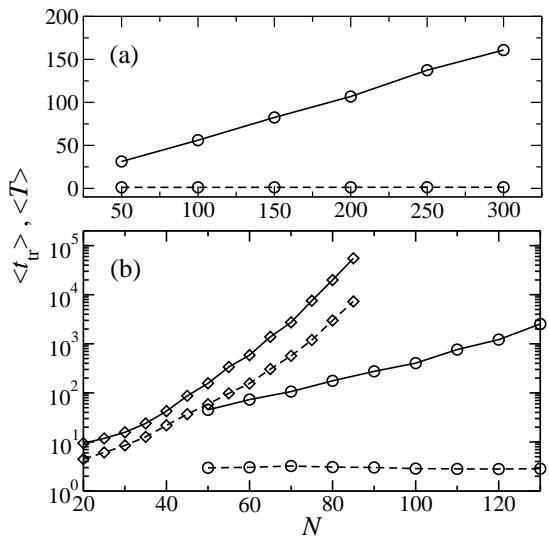}
  \caption{Average length 
  $\av{t_{\rm tr}}$ (solid line) of transient and
  corresponding period $\av{T}$
  of the periodic attractor (dashed line) {\em vs.} system size:
  (a) $G=0.5\,$; (b) $G=1.8$ (circles) and $G=2.5$
  (diamonds). Note in (b) the logarithmic scale of the vertical axis.}
  \label{fig:trans2}
\end{figure}

A more precise characterization of the dynamics is obtained by investigating
the scaling behavior with $N$. In Fig.~\ref{fig:trans2}a, we see that for $G=0.5$,
the transient length increases linearly  with the system size $N$, while the
average period $\langle T\rangle$  of the asymptotic attractor remains almost 
constant. At the same time, in Fig.~\ref{fig:trans2}b, we see that for $G=1.8$ and
$G=2.5$, the transient grows exponentially fast. Finally, it turns out that
for $G=2.5$ also the average period grows exponentially. We consider this
as a preliminary indication that at larger coupling strength another transition
may occur. However, here we focus our attention on the qualitative changes
occurring for $G\approx 1$ where only the transient starts growing
exponentially.

Another important point to be carefully investigated are the statistical 
properties of the transient dynamics. In particular, we start from
the stationarity that can be analysed by computing the the so--called
coefficient of variation ($CV$), that is determined by subdividing a time
series $x(n)\,,\ n=1,2,\dots, L\,$, into a sequence of windows of constant
duration $l$, and then computing
\begin{gather*}
{\bar x}(m) = \frac{1}{l}\sum_{i=(m-1)l +1}^{ml} x(i)\,,\ m=1,2,\dots, L/l\ , 
\\
\sigma(m) = \left[\frac{1}{l}\sum_{i=(m-1)l +1}^{ml} (x(i) - 
{\bar x}(m))^2\right]^{1/2}\ .
\end{gather*}
The coefficient of variation is thereby defined as
\be
CV(m) = \sigma(m)/{\bar x}(m)\ .
\ee
The parameter $l$ must be chosen in such a way that each window contains a
sufficient number of samples to reliably compute the local average and the
standard deviation.

\begin{figure}[h]
\includegraphics[width=0.4\textwidth,clip=true]{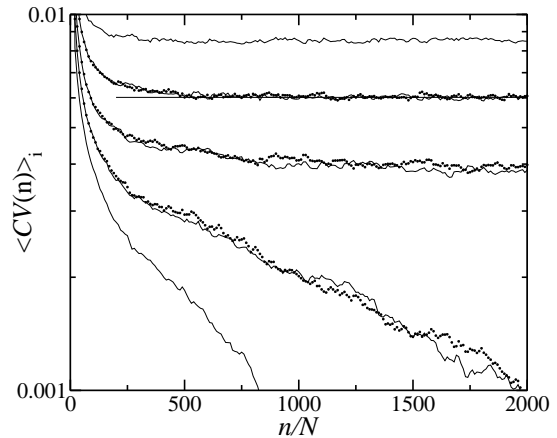}
  \caption{Solid lines: Coefficient of variation as function of integer time 
  for $N=1000$ and (in ascending order starting from the bottom)
   $G=0.9,1.0,1.1,1.2,1.3\,$. Stars: The same for a different disorder 
realization and $G=1.0,1.1,1.2\,$.
  The straight line corresponds to the saturation value at $G=1.2\,$.}
  \label{fig:isivar1}
\end{figure}

In our case, the time series of interest is the sequence of single neuron
interspike intervals $T$, obtained by iterating map (\ref{eq:map2}).
All intervals are recorded over a time window of length $l=10N$, so that each 
neuron fires on average ten times. In order to reduce statistical fluctuations,
the $CV$ has been further averaged over 100 different initial conditions for a
fixed network realization. In Fig.~\ref{fig:isivar1}, we have plotted the time
evolution of $\av{CV(n)}_{\rm i}\,$ for different values of $G$. Moreover, for
each value of $G$, the behavior of $\av{CV(n)}_{\rm i}\,$ is presented for two
different network realizations to convince the reader that a further averaging
on the disorder is not truly required to conclude about the stationarity of the
regime. Indeed, the main point is that for $G<1$ the transient is nonstationary,
as $\av{CV(n)}_{\rm i}$ vanishes exponentially while approaching the periodic
attractor. On the other hand, for $G\agt 1.2$, $\av{CV(n)}_{\rm i}$ approaches a
finite value, thus confirming that the ``transient" regime is stationary and
one can thereby meaningfully speak of an invariant measure and pose the
question of determining its correlation properties.

We have also performed other stationarity tests based, e.g., on the analysis of
distributions of temporal distances of neighboring points in state space (e.g.,
see \cite{Riecke-Sternickel:2002}). They confirm the stationarity of the
transients above the transition and confirm the existence of a transition
point in the interval $1.2<G<1.3\,$. As these tools do not provide further
insight we do not further comment on their outcome.

\begin{figure}[h]
\includegraphics[width=0.42\textwidth,clip=true]{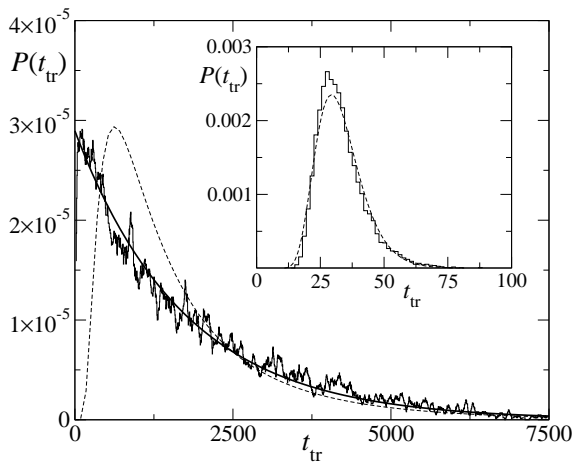}
  \caption{Normalized histogram (random initial conditions at fixed disorder, 
  $N=65$) of transient lengths for $G=2.1\,$. The solid curve is an exponential,
  the dashed curve an inverse Gaussian with equal mean and variance. The inset
  shows the distribution with an inverse Gaussian fit for $G=1\,$.
  }\label{fig:dist2}
\end{figure}

The transient length strongly depends on the initial condition. In
Fig.~\ref{fig:dist2} we show the probability distribution function
$P(t_{\rm tr})$ for a given realization of the disorder in both phases.
For small $G$,  $P(t_{\rm tr})$ can be confidently fitted with an inverse
Gaussian (see the inset of Fig.~\ref{fig:dist2}), which is the typical
statistical distribution of biased escape--time problems
\cite{Chikara-Folks:1988}. This confirms once more the nonstationarity of the 
corresponding transient dynamics, since it is driven by a systematic drift
towards the asymptotic periodic state. Above the transition, $P(t_{\rm tr})$ is 
Poissonian: the transient statistics can be therefore interpreted as a typical
escape process through an activation barrier from a locally equilibrated system.
Again, this evidence supports the conjectured stationarity of the transient
dynamics for $G\agt 1$ (in the large $N$ limit).

Finally, $P(t_{\rm tr})$ does not change qualitatively for different
realizations of disorder, although sizeable quantitative differences can be
found for small values of $N$. However, our simulations indicate that the
probability distribution $P(t_{\rm tr})$ is a self-averaging quantity since, 
upon increasing $N$, sample-to-sample fluctuations become increasingly small.
This is absolutely clear for $G \approx <1$, when it is possible to compute
the transient time even for moderately large values of
$N \sim {\cal O} (10^4)$. Above $G=1$, where direct simulations are almost
undoable, there is, however, no reason to expect a qualitative change,
given all the evidence of an ergodic dynamics.

Altogether, the ``transient" dynamics observed for $G\approx >1$: (i) is
characterized by a negative Lyapunov exponent; (ii) is effectively stationary;
(iii) lasts for exponentially long (with the system size) times. These are the
distinguishing properties of ``stable chaos", a phenomenon extensively
investigated in coupled map lattices \cite{stable}, where similar features of
the transient statistics have been observed \cite{Livi-Mekler-Ruffo:1990}.
However, it is worth recalling that stable chaos has been observed also in
chains of forced oscillators \cite{bonaccini} and recently uncovered in the 1D
hard point gas of diatomic particles \cite{cipriani}.
 
In all such instances, stable chaos is associated with the presence of
discontinuities in phase space. In the coupled map models, the discontinuities
are transparent in the definition of the local piece-wise linear maps
\cite{stable}. In the hard-point gas they arise in connection to three-body
collisions. Around any configuration leading to one such multiple collision,
the ordering of the two-body collisions changes abruptly. The non-commutativity
of the collision itself induces therefore a discontinuity in the dynamics
that is conceptually similar to those postulated in the coupled-map lattices.
The same happens in the present context. In fact, let us consider the
time-evolved one-parameter family of configurations,
${\bf v}^\nu(t) = \{ v_1^\nu(t), v_2^\nu(t), \ldots, v_{k}^\nu(t), v_{k+1}^\nu (t), \ldots, v_{N}^\nu (t) \}$ 
where
$v_j^\nu(0)= u_j +\nu\delta_{jk}$ and $\delta_{jk}$ is the Kroenecker
delta function. It is easy to convince oneself that ${\bf v}_\nu(t)$ 
breaks into two disconnected parts when $v_k^\nu(t)=1$ if $\nu$ is such that
$v_k^\nu(t)=v_{k+1}^\nu(t)$ and there is only one connection between neurons
$k$ and $k+1$. The discontinuity is due to the fact that only one of them
inhibits the other. It is therefore interesting to verify that, in agreement
with the past observations, the presence of discontinuities in the phase-space
is a condition for the onset of a ``stable chaos" dynamics. If no anomalous
transients arise for small coupling strengths, it is because the condition is
necessary by not at all sufficient.

\section{Two dynamical phases}
\label{five}
\begin{figure}[h]
\includegraphics[width=0.36\textwidth,clip=true]{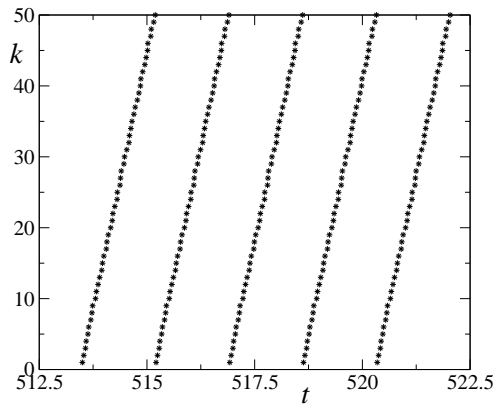}
\caption{ Firing pattern (index of the firing neuron {\em vs.} time) 
of a typical periodic attractor
for $G=2,N=50\,$ in the LP; $k$ is the index of the firing neuron.}
\label{fig:perpat1}
\end{figure}

In this section we study the properties of the dynamical phases and perform a
preliminary analysis of the transition. For $G\alt 1$, the evolution is similar
to that of the fully coupled case. After a short transient time the system
converges to a state characterized by a sequence of $N$ spikes (emitted by the
$N$ neurons), which periodically repeats itself (see Fig.~\ref{fig:perpat1}). All
neurons fire with the same pace ($T(m) = {\rm const.}$), but their phases are no
longer equispaced (i.e. $t_{\rm ISI}(n)$ varies in time). In other words, the
asymptotic solution is a phase--locked, i.e. synchronized, state. In the
following, this dynamical regime will be denoted {\it Locked Phase} (LP). The
main difference with the fully coupled case is that different spike sequences
are not equivalent to one another. In the limit of identically-coupled neurons
the dynamics is invariant under any permutation, but this is no longer true as
soon as some degree of heterogeneity is introduced in the network connectivity.
As a result, for $r_m \agt 0$ there exists an exponentially large number of
different periodic attractors. More precisely, the total number of attractors
can be estimated to be of the order of $N !/ (\prod_{d=1}^M N_d !)$, where $N_d$ is 
the number of neurons connected to the same (incoming and outgoing) neurons in
the network and $M$ is the number of these equivalence classes present in a
given network realization.

\begin{figure}[h]
\includegraphics[width=0.38\textwidth,clip=true]{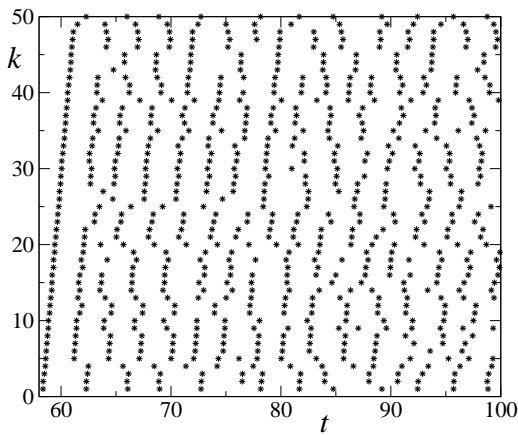}
\caption{
Firing pattern for $G=2,N=50\,$ in the UP; $k$ is the index of the firing neuron.}
\label{fig:pattern}
\end{figure}

For larger coupling strengths, an irregular dynamical regime arises that we call
{\it Unlocked Phase} (UP), because the mutual ordering keeps changing during the
entire transient, as one can, e.g., see in Fig.~\ref{fig:pattern}, where slow but
systematic pattern adjustments can be seen in the whole time range.

\begin{figure}[h]
\includegraphics[width=0.42\textwidth,clip=true]{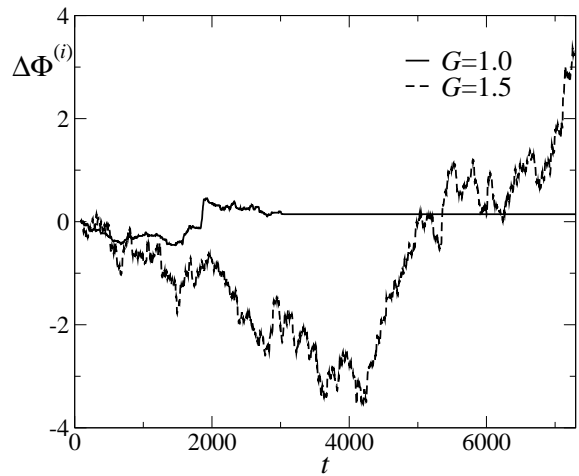}
\caption{Time evolution of the local phase difference $\Delta \Phi^{(i)}(t)$
for $G=1.0\,$ (solid line) and $G=1.5\,$ (dashed line). We consider
a system with $N=1000$ and diluition $r_{m}=0.05\,$.}
\label{fig:phase}
\end{figure}

One can better elucidate this dynamical phase by looking at the relative
potential differences. Without prejudice of generality, we choose one of the
$N$ neurons, say neuron $i$, and consider the subsequence of its firing events.
Given the spike emitted at time $t$, we compute the difference between the last
ISI, $T^{(i)}$, of that neuron and the average ISI of the $j$ neurons firing
immediately before and after time $t$. We denote this time--dependent quantity
by $\Delta \phi^{(i)}(t)$. The value of $j$ has to be chosen neither too large,
to avoid inclusion of spikes from the same neurons, nor too small, to get rid
of local fluctuations. We have checked that the choice $j=5$ is a reasonable
compromise. The integrated phase--shift
$\Delta \Phi^{(i)}(t) = \int_0^t dt'\Delta\phi^{(i)}(t')$  is plotted in
Fig.~\ref{fig:phase} for a generic initial condition and a typical realization
of the disorder. In the LP, after a short transient, $\Delta \Phi^{(i)}$
collapses onto a constant value, which indicates that the system has converged
to a periodic firing sequence with the neurons firing with the same pace.
On the other hand, in the UP, $\Delta \Phi^{(i)}$ wanders erratically, 
performing a sort of unbiased random--walk. We can indeed confirm that at
least for $G \alt 1.8$ and within statistical fluctuations all neurons are
characterized by the same average firing rate. 

\begin{figure}[h]
\includegraphics[width=0.36\textwidth,clip=true]{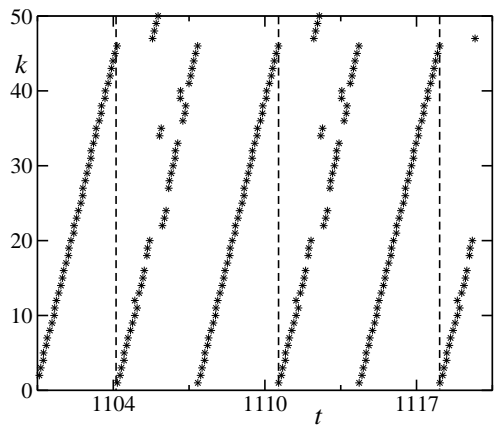}
  \caption{Firing pattern associated with a periodic attractor in the UP 
for $G=2.0, N=50\,$. The period is indicated by the vertical lines.}
\label{fig:perpat2}
\end{figure}

A spontaneous symmetry breaking induced by disorder appears at larger
coupling strengths when we can, e.g., find periodic orbits with different
neurons exhibiting different firing rates (see Fig.~\ref{fig:perpat2}).
This phenomenon is, however, not connected with the onset of exponentially long
transients that appears for significantly smaller $G$ values.

\begin{figure}[h]
\includegraphics[width=6truecm,height=6truecm,clip=true]{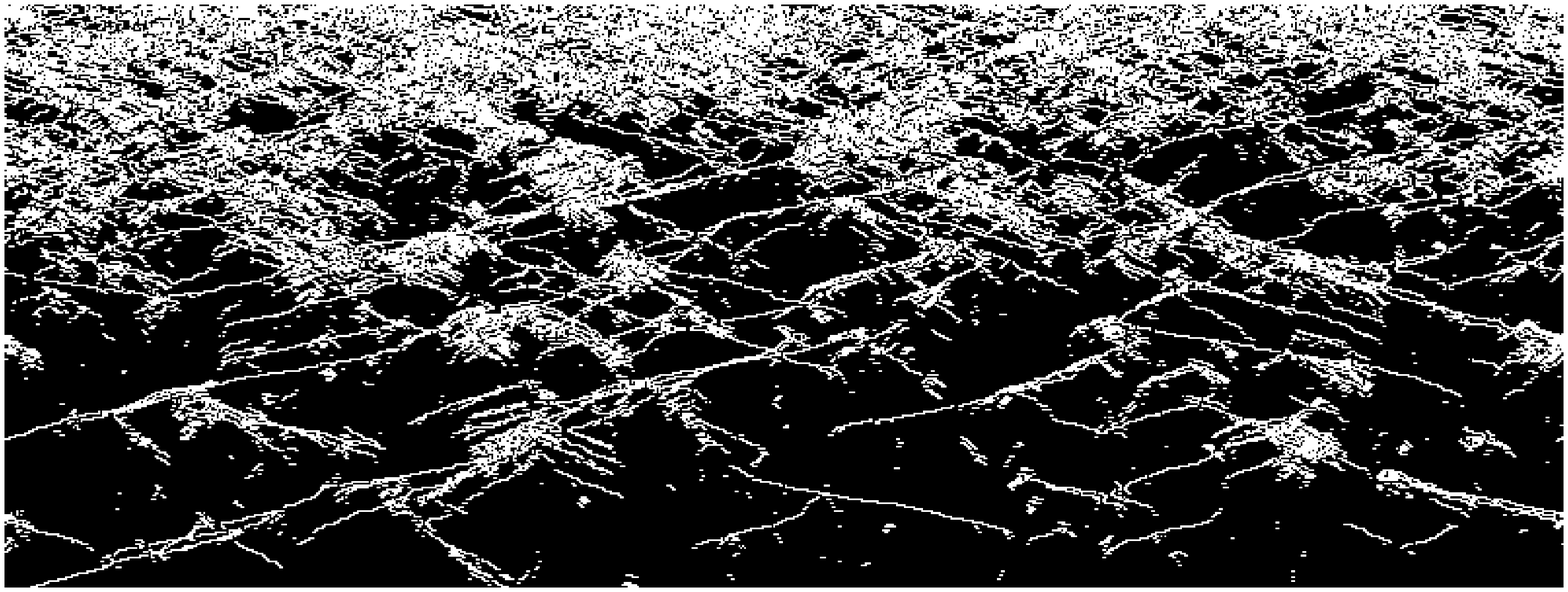}
\includegraphics[width=6truecm,height=6truecm,clip=true]{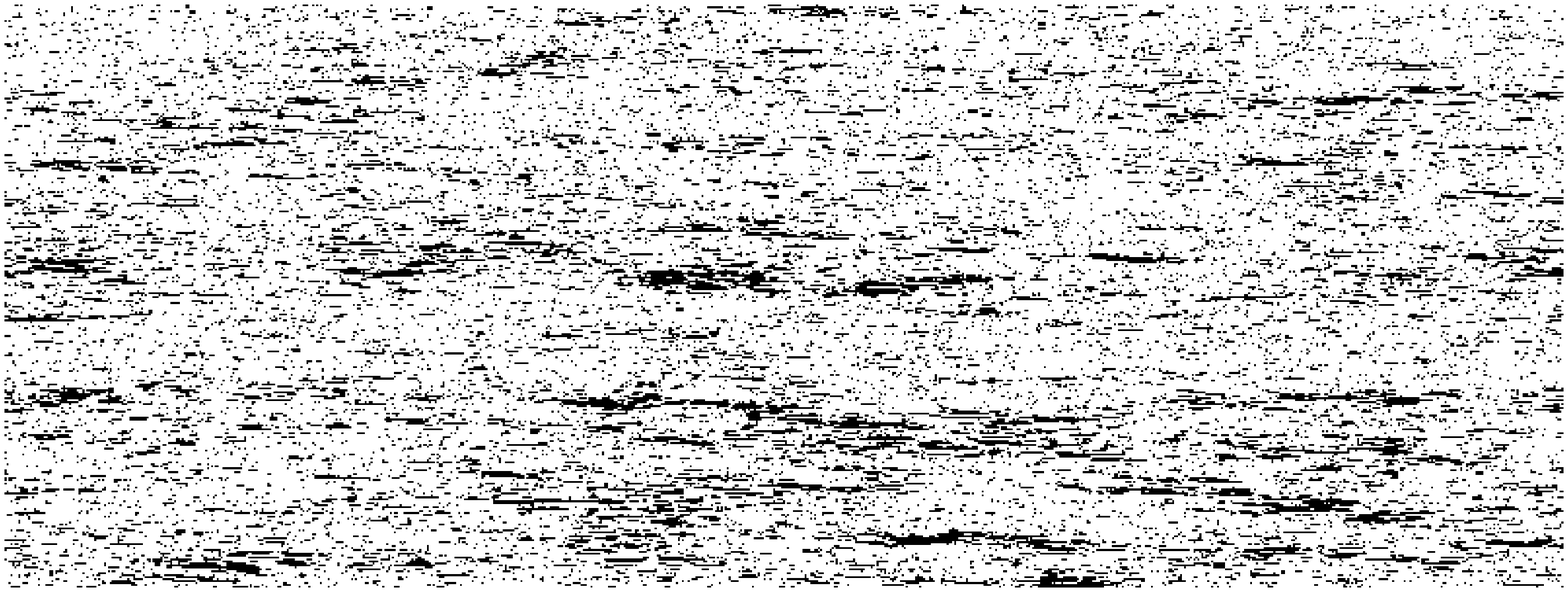}
\caption{Patterns associated with the two different phases of the dynamical
system. On the vertical axis (from top to down) is reported the
time, while the horizontal axis represents the index of successive firing
neurons. The upper pattern refers to LP for $G=0.5\,$ , and the lower 
to the transition region with $G=1.3\,$. In both cases a
system with $N=801$ and dilution $r_{m}=0.05$ is considered.}
\label{fig:greyplot}
\end{figure}

Further information about the two dynamical phases can be gained 
from the introduction of a suitable space-time representation. 
Having verified that all neurons statistically behave in the same manner, we
can split the entire spike series into sequences of $N$ consecutive events and
label the sequences with an index $m$, which naturally plays the role of time.
On the other hand, the index $i$, labelling the position within a single
sequence, plays the role of space. Finally, we introduce a binary variable
$b(m,i)$ to distinguish two cases: whenever the neuron emitting the $i$-th spike
in the $m$-th sequence is the same neuron which emits the $i$-th spike in the
$(m-1)$-st sequence, we set $b(m,i)=0$; otherwise $b(m,i)=1$. The resulting
patterns are reported in Fig.~\ref{fig:greyplot}, where 0, 1 are coded as black
and white colours, respectively. In the UP, the pattern is very irregular and
no order seems to emerge even after a very long time lapse. Conversely, in the
LP, black tends to prevail quite soon, indicating that the system rapidly
converges to a fixed firing sequence, i.e., to a periodic attractor. The most
interesting features of this representation are the white {\it defects}
propagating in the black background. They indicate that the firing rate of some
neurons is temporarily slower or faster than that of the ``neighbouring''
ones. These patterns are strongly reminiscent of a Directed Percolation (DP)
transition. Actually, defects tend to be eliminated in the LP, which is
analogous to the absorbing phase of DP. On its turn, the UP shows similarities
with the so called ``active" phase of DP, ruled by a persistent defect dynamics.
It is remarkable that the dynamics of an almost globally coupled system
exhibits distinctive features of one-dimensional systems with short-range
interactions, as it is the case of 1+1 contact processes. The analogy appears
less astonishing after recalling that the evolution of the fully coupled model
is described by a suitable partial differential equation (see Eq.~(\ref{eq:12})).
However, including the role of disorder and proving a possible analogy with
directed percolation is not an easy task. We prefer to leave the elucidation of
this problem to future analyses and here we concentrate our efforts to
obtaining a refined characterization of the two dynamical phases. 

\begin{figure}[h]
\centerline{\includegraphics[width=0.42\textwidth,clip=true]{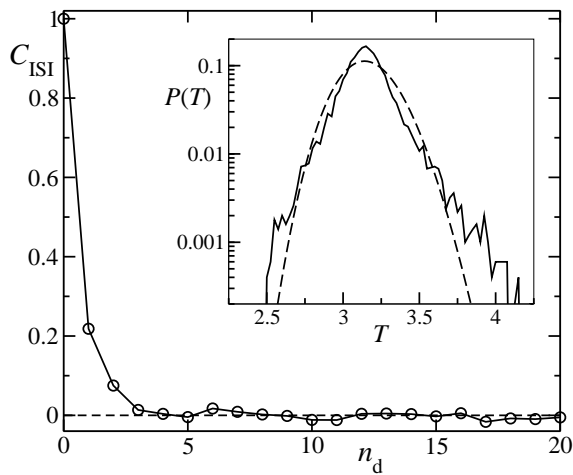}}
\caption{The normalized autocorrelation $C_{\rm ISI}$ of the $T^{(i)}$ during
transient for a single neuron i {\em vs.} the integer delay for $N=1000 ,
G=2.0\,$. The inset shows the corresponding distribution of the $T^{(i)}$ and an
inverse Gaussian with the same mean and variance.}\label{fig:isicorr}
\end{figure}

In particular, we have determined the distribution and the time autocorrelation
function of the single-neuron interspike time interval $T^{(i)}(m)$. In the UP,
the normalized autocorrelation function $C_{\rm ISI}$ is plotted in
Fig.~\ref{fig:isicorr}. It exhibits an exponentially fast decay, typical of both
chaotic and stochastic regimes: memory is lost after typically 3 to 4 firing
events. This quantitative confirmation of the ``irregularity" is yet another
element strengthening the analogy with stable chaos. The ``stochastic"-like
character of the evolution is also confirmed by the shape of the Probability
Density Function (PDF), $P(T^{(i)})\,$, which has a Poisson-type tail but is also
not much different from an inverse Gaussian -- a typical distribution
encountered in neural dynamics \cite{tuckwell}.

Finally, given the evidence of the effective stochastic behaviour, we develop a
standard mean-field approach to characterize the behaviour of the network in the
large $N$-limit. In particular, one can conveniently approximate the mean field
with a deterministic drift plus a zero-average stochastic process. This amounts
to replacing the initial model with the following stochastic equation (to be
interpreted in the Ito sense, as the noise arises from a discrete-time
equation),
\begin{equation}
\label{eq:mf0}
\dot{v}_j=c-v_j - \frac{G(v_j+w)}{T} + \frac{G(v_j +w)}{(1-r_{m})}\, \eta_j(t)
\,\,\, , \,\,\, v\in [0,1)\ ,
\end{equation}
where the time $T$ is determined self-consistently, while $\eta_j(t)$ is a
delta-correlated stochastic process 

$$\langle \eta_i(0) \eta_j(t) \rangle = \frac{r_m(1-r_m)}{TN} \delta_{ij}\delta(t)\, ,$$

with $\delta_{ij}$ and $\delta(t)$ being the Kroenecker and Dirac $\delta$-functions,
respectively. In the thermodynamic limit the noise term vanishes and can
therefore be neglected in the computation of average properties. In particular,
by integrating the deterministic force field in Eq.~(\ref{eq:mf0}), one obtains
an  implicit equation for $T(G)$,
\be
\label{tav}
\nonumber
G=\frac{x}{1+x}\ln{\frac{c-xw}{c-1-x(1+w)}} \quad,
\ee
where $x=G/T(G)$. This formula holds not only in the UP, but also in the LP,
because the main difference between the two regimes does not concern the noise 
amplitude, but its correlation properties. In fact, the relative diffusion of 
the generic neurons $i$ and $j$ depends on the fact that the stochastic 
signals $\eta_i$ and $\eta_j$ are not correlated with one another. If this is 
the case, the relative diffusion induced by the noise over a time equal to the 
average ISI is $1/\sqrt{N}$. Since, on the other hand, the average
distance between consecutive neuron potentials is on the order of $1/N$,
we conclude that for any finite $G$ and large enough $N$ the diffusion is
sufficiently strong as to scramble the neurons. The smallness of the noise
amplitude also explains why
at a coarse-grained level the firing patterns appear almost periodic (see
e.g. Fig. \ref{fig:pattern}).
Now, in order to complete a
self-consistency argument, it would be necessary to connect the changes in the 
neuron ordering with the correlation properties of the noise terms in 
Eq.~(\ref{eq:mf0}). Since different neurons have different connection trees,
$\eta_i(t+T)$ will be decorrelated from $\eta_i(t)$ and, more important, 
$\eta_i(t+T)-\eta_j(t+T)$ will change with respect to the previous ISI.
However, transforming these hand-waving arguments into a quantitative criterion
for the identification of the critical value $G_c$, above which self-generated
fluctuations can be robustly sustained, is not an easy task and we leave it to 
future studies. Here, we limit ourselves to stressing the relevant role played 
by correlations between neighbouring neurons (i.e. those with the closest
--to--threshold  membrane potentials). Below $G_c$ the neurons rearrangement
continues until it stops when a suitable ordering is reached, characterized by
the property that neighbouring neurons do not diffuse away from each other. The
existence of exponentially long transients in the UP tells us that such an
arrangement does not exist in the strong coupling regime.

The two-phase scenario here above described persists for a wide range of $r_m$
values. While no qualitative changes are expected to appear until $r_m$ becomes
so large that the network is decomposed into uncoupled blocks, we are unable to
formulate any conjecture for the limit case $r_m \to 0$, due to the need of
studying very large networks.

\section{Conclusions and open problems}
\label{six}

The main result of this paper is the observation that a network of leaky
integrate-and-fire neurons can exhibit a pseudo-chaotic behaviour in spite of
an entirely negative Lyapunov spectrum. Previous studies\cite{bonaccini} have
identified in strong localized nonlinearities a necessary (and far from
sufficient) condition for this phenomenon to exist. Discontinuities are indeed
responsible for a sudden amplification of (finite-amplitude) perturbations and
for the resulting irregular dynamics. In the models where this phenomenon was
initially observed, the discontinuities are somehow artificial features of the
local maps, but it has been recognized that they may also spontaneously emerge.
This is the case of the hard-point diatomic gas, where discontinuities arise in
the vicinity of three-body collisions\cite{cipriani}. Here, we have seen that
they also occur in neuronal networks, in connection with changes in the spike
ordering. However, when and why a discontinuity may be so important as to
steadily sustain (in the thermodynamic limit) an irregular dynamics is still a
completely open problem.

Another open question concerns the nature of the phase-transition.
Macroscopically, it manifests itself as a collective de-synchronization, but
there are difficulties in quantitatively describing the phenomenon. On one
hand, it cannot be analysed by looking at the linear stability of some solution,
since linear stability is ensured above and below the transition. Moreover, the
absence of local instabilities (standard deterministic chaos) makes it
questionable the concept of a transition point. In fact, the study of a partially
analogous transition in a coupled map lattice revealed that regular and
irregular phases are separated by a finite fuzzy region, rather than by
a point-like transition \cite{Cecconi-Livi:1998}. Perhaps the almost global
coupling of the network studied in this paper may induce a standard transition
scenario and the analogies with contact processes suggested by
Fig.~\ref{fig:phase} indicate that this is a reasonable expectation. However,
for the moment, we must limit ourselves to observing that the scenario is 
robust against various modifications, such as the addition of further disorder
accounting for a non homogeneous strength of the coupling and specific
modifications of the network topology.

Recently, it has been pointed out that an irregular dynamics should
significantly enhance information processing \cite{Masuda-Aihara:2002}. At
variance with sparsely coupled chaotic networks, where chaos results from the
balance of inhibition and excitation \cite{Vreeswijk-Sompolinsky:1996}, we have
found irregular firing patterns, characterized by a Poisson-like distribution
of interspike intervals, in the presence of linear stability and absence of
excitation. It would be worth investigating if and how such a Lyapunov stable
irregular dynamics is an effective tool for information processing.
In the literature there is a growing evidence of recurrent motifs in the
firing pattern (see e.g. \cite{Zhigulin}), which suggests that phase lockings may
be a tool to encode information. In our model, recurrent motifs appear
naturally in the UP and yet keep evolving, indicating a sort of information
processing which certainly deserves a thorough investigation from this
point of view.

\acknowledgments

We thank T.~Kreuz for performing the more elaborate stationarity tests. 
We further acknowledge MIUR-PRIN05 for partial support
within the project on "Dynamics and thermodynamics of
systems with long range interactions".



\begin{thebibliography}{9}
\bibitem{foot0} Everywhere in this paper, we exclude the zero Lyapunov exponent
corresponding to perturbations oriented along the trajectory itself. This
will be automatically done, when, in the following, we implement a
discrete-time mapping, by introducing a suitable Poincar\'e section.
\end{thebibliography}
\end{document}